\documentclass[showpacs,preprintnumbers,amsmath,amssymb]{revtex4}
\makeatletter
\begin{document}
\title{NONCOMMUTATIVE GEOMETRY AND GEOMETRIC PHASES}
\author{B. Basu}
 \email{banasri@isical.ac.in}
\author{Subir Ghosh}
 \email{subir_ghosh2@rediffmail.com}
 \affiliation{Physics and Applied Mathematics Unit\\
 Indian Statistical Institute\\
 Kolkata-700108 }
 \author{S. Dhar}
\email{sarmi_30@rediffmail.com} \affiliation{S. A. Jaipuria
College\\Kolkata- 700005 }

\begin{abstract}
We have studied particle motion in generalized forms of
noncommutative phase space, that simulate  monopole and other
forms of Berry curvature, that can be identified as effective
internal magnetic fields, in coordinate and momentum space. The
Ahranov-Bohm effect has been considered in this form of phase
space, with operatorial structures of noncommutativity. Physical
significance of our results are also discussed.
\end{abstract}
\maketitle
\section{Introduction}
 The possible existence of magnetic monopole (MM) was first discussed by Dirac
\cite{dirac}and later in \cite{thooft}in non-abelian gauge theory. 

However, recently \cite{mono1} the signatures of MM in {\it{crystal momentum
space}} in {\bf{SrRuO}}$_ {\bf 3}$, (a ferromagnetic crystal),
have appeared as peaks in transverse conductivity $\sigma_{xy}$.
The MM formation in low energy regime ($\sim 0.1-1$eV) in the
condensed matter system \cite{mono1}, (as compared to the
predicted range $\sim 10^{16}$GeV in particle physics
\cite{thooft}), is obviously the reason for their observation in
the former. The MM in $\sigma_{xy}$ is again directly linked to
Anomalous Hall Effect (AHE) where $\sigma_{xy}$ is identified as
the Berry curvature. The very intrinsic origin of AHE \cite{lutt},
independent of external magnetic fields, suggests \cite{berard}
that the whole phenomena might be interpreted as a motion of
(Bloch) electrons in a non-trivial symplectic manifold with the
symplectic two-form being essentially the Berry curvature.
This is  a specific form
of Non-Commutative (NC) space (see for example \cite{nc}), that
appears because one introduces \cite{niu,mono1} a gauge covariant
position operator 
$x_\mu  \equiv i\partial_{\mu}-a_{n\mu}(\vec
k),~a_{n\mu}=i<u_{n\vec k}\mid \frac{\partial}{\partial
k_{\mu}}u_{n\vec k}>$, 
for the Bloch wavefunction $\psi_{n\vec k}(r)=exp(i\vec k.\vec
r)u_{n\vec k}(r)$, with the coordinates satisfying the NC
structure,
$[x_\mu ,x_\nu
]=-iF_{\mu\nu},~~F_{\mu\nu}=\partial_{k_{\mu}}a_{n\nu}-\partial_{k_{\nu}}a_{n\mu}$.
 This NC induces the additional
anomalous part of velocity that yields the AHE. Notice that for
the crystalline system in question, Bloch states are the natural
setting and the curvatures become functions of momenta as
expected. Clearly $a_{n\mu}$ is unrelated to any external source
and is generated within.

In a
recent paper \cite{pur} real space Berry phase manifests as a
further contribution   in AHE in {\bf{AuFe}} alloy and the
underlying theory \cite{kawa} requires a topologically non-trivial
spin configuration. The theory \cite{kawa} indicates that the
coupling between this net spin chirality and a global
magnetization, (which might be spontaneous, as in {\bf{AuFe}}
alloy), plays a crucial role. Onoda, Tatara and Nagaosa in
\cite{kawa} have argued that the real space and momentum space
Berry phase manifest itself in two different regimes. The real
space vortex  is a good picture in disordered case (equivalently
for short electrons with  mean free path) whereas momentum space
vortex  is a useful model in the pure case \cite{gen1}. A
re-entrant ferromagnet, such as the {\bf{AuFe}} alloy is a sample
of the former kind. It should also be remembered that \cite{gen1}
in principle, complicated structures of Berry curvature are indeed
possible depending on the particular nature of a sample, although
so far the only numerical work concerns the simple monopole form,
as observed in \cite{mono1}. Besides, a study of the underlying
geometry of the ferromagnetic spin system shows that the inherent
magnetic type of behavior is caused by the Berry curvature in real
space
 which arises due to the spin rotations of conducting electrons and is the effect of  noncommutativity in momentum space \cite{bb1}.

Keeping this background in mind, we put forward forms of NC space
that can induce singular behavior (in the effective magnetic
field)  in coordinate space. Different novel structures of Berry
curvature appear in our framework. Incidentally, our work is a
generalization of the work of \cite{berard}. The NC structure and
its associated symplectic form, considered in \cite{berard}, was
not general enough to allow the vortex structures that we have
obtained here.

With this special form of NC space we have calculated the

Aharanov- Bohm (AB) phase and have shown that there is a
modification term due to the noncommutativity of space-space
coordinates. This leads to new expression and bound for $\theta $
- the noncommutativity parameter.

The paper is organized as follows: In Section {\bf{II}} we
introduce the particular form of NC space that will be studied
subsequently. Section {\bf{III}} deals with the Lagrangian
formulation of the model and the related dynamics in a general
way. Section {\bf{IV}} is devoted to the study of the
Aharanov-Bohm effect in this specific NC phase space . In Section
{\bf{V}} we discuss the physical implications of our findings.
\section{Noncommutative phase space}
We start by positing a non-canonical phase space that has the
Snyder form of spatial noncommutativity and at the same time the
momenta satisfies a conventional monopole algebra. Similar
structures have also appeared in \cite{lages}. In the beginning we
have introduced two distinct NC parameters $\theta$ and $b$ for
the above two independent forms of noncommutativity so that their
individual roles can be observed.

The phase space is given below:
\begin{eqnarray}\label{aa1}
\left \{X_i, X_j\right\}&=&  -\theta(X_i P_j - X_j P_i),\nonumber\\
\left\{X_i, P_j\right\}&=&\delta_{ij}  -\theta P_i P_j,\nonumber\\
\left\{P_i, P_j\right\}&=& b \epsilon_{ijk} \frac{X_k}{X^3},
\end{eqnarray}
where $X=\sqrt{X^2}$.

We discuss rotational properties of the vectors.
 From the definition of the angular momentum, $L_j=\epsilon_{jkl} X_k P_l$ we
have the following commutation  relations,
\begin{eqnarray}\label{ll1}
\left\{X_i,L_j\right\}& =& \epsilon_{ijk} X_k,\nonumber\\
\left\{P_i,L_j\right\} &= &\epsilon_{ijk} P_k +
b \left(\frac{X_i X_j}{X^3} ~-~\delta_{ij}\frac{1}{X}\right),\nonumber\\
\left\{L_i,L_j\right\} &=& \epsilon_{ijk} L_k + b
\epsilon_{ijk}\frac{X_k}{X}.
\end{eqnarray}
Notice  that $b$ (and not $\theta$) destroys the transformation
properties. But this is expected since as is well known the Snyder
algebra does not clash with rotational invariance.   To restore
the angular momentum algebra consider the
term\begin{equation}\label{mm1}
    S_j=-d \frac{X_j}{X}
\end{equation}
which yields the total angular momentum as
\begin{equation}\label{ll2}
    \tilde{L}_j = L_j+S_j.
\end{equation}
We naturally identify $\vec S$ as the effective spin vector, that
is induced by the algebra (\ref{aa1}).  Now angular momentum
algebra is given by
\begin{eqnarray}\label{mm2}
  \{X_i,\tilde{L}_j\} &=& \epsilon_{ijk} X_k  -\theta d \left(-\frac{X_i P_j}{X}+\frac{X_i X_j(X.P)}{X^3}\right)\nonumber\\
  \left\{ P_i,\tilde{L}_j\right\} &=& \epsilon_{ijk} P_k
         +(b-d)\left(\frac{X_i X_j}{X^3} ~-~\delta_{ij}\frac{1}{X}\right)
         -\theta d \left(\frac{P_i P_j}{X}-\frac{P_i X_j(X.P)}{X^3}\right) \nonumber\\
  \left\{ \tilde{L}_i,\tilde{L}_j\right\} &=& \epsilon_{ijk}
  \tilde{L}_k +(b-d)
\epsilon_{ijk}\frac{X_k}{X}
\end{eqnarray}

Putting $d=b$ the algebra is as follows
\begin{eqnarray}\label{mm3}
  \{X_i,\tilde{L}_j\} &=& \epsilon_{ijk} X_k  -\theta b \left(-\frac{X_i P_j}{X}+\frac{X_i X_j(X.P)}{X^3}\right)\nonumber\\
  \left\{ P_i,\tilde{L}_j\right\} &=& \epsilon_{ijk} P_k
  -\theta b \left(\frac{P_i P_j}{X}-\frac{P_i X_j(X.P)}{X^3}\right) \nonumber\\
  \left\{ \tilde{L}_i,\tilde{L}_j\right\} &=& \epsilon_{ijk}
  \tilde{L}_k
\end{eqnarray}
The above considerations prompt us to study a simpler NC algebra,
with $b=\theta $,
\begin{eqnarray}\label{a1}
\left\{X_i, X_j\right\}&=& -\theta (X_i P_j - X_j P_i),\nonumber\\
\left\{X_i, P_j\right\}&=&\delta_{ij}  -\theta P_i P_j,\nonumber\\
\left\{P_i, P_j\right\}&=& \theta \epsilon_{ijk} \frac{X_k}{X^3},
\end{eqnarray}
Hence the NC algebra is governed by a single NC parameter
$\theta$.

Hence to the approximation ({\it{i.e.}} $O(\theta )$) that we are
interested in and ignoring the terms which are in order of
$\theta^2$, we find,
\begin{eqnarray}\label{m2}
  \{X_i,\tilde{L}_j\} &=& \epsilon_{ijk} X_k, \nonumber\\
  \left\{ P_i,\tilde{L}_j\right\} &=& \epsilon_{ijk} P_k , \nonumber\\
  \left\{ \tilde{L}_i,\tilde{L}_j\right\} &=& \epsilon_{ijk}
  \tilde{L}_k.
\end{eqnarray}
Thus, to the lowest non-trivial order in $\theta$, is possible to
define an angular momentum operator in a consistent way.

Since the form of noncommutativity is operatorial in nature we
must check that the Jacobi identities are satisfied,
$$ J(A,B,C)\equiv \left\{A,\{B,C\}\right\}+ \left\{B,\{C,A\}\right\}+\left\{C,\{A,B\}\right\}=
0,$$ We obtain,
\begin{equation}\label{j1}
J(X_i,X_j,X_k)=J(X_i,X_j,P_k)=J(X_i,P_j,P_k)=0.
\end{equation}
The Jacobi identity between transformed angular momentum is also
satisfied,
\begin{equation}\label{j3}
 J( \tilde{L}_i,\tilde{L}_j,\tilde{L}_k)=0.
\end{equation}
However the  Jacobi identity between momenta is violated,
\begin{eqnarray}\label{j2}
J(P_i,P_j,P_k)=
-3\theta\epsilon_{ijk}\frac{1}{X^3}+3\theta\frac{X_l}{X^5}
   \left(\epsilon_{ijl}X_k+\epsilon_{jkl}X_i+\epsilon_{kil}X_j\right)\nonumber \\
   ~~~~~~~~~~= -\theta \epsilon_{ijl}\partial_l \left(\frac{X_k}{X^3}\right)
         -\theta\epsilon_{jkl}\partial_l \left(\frac{X_i}{X^3}\right)
         -\theta\epsilon_{kil}\partial_l
         \left(\frac{X_j}{X^3}\right).
\end{eqnarray}
But from the analysis of Jackiw \cite{dirac} we know the
implications of this violation and how to live with it.
\section{Symplectic Dynamics}
Non-violation of the Jacobi identities (at least up to the
prescribed order) is essential in our case since we wish to study
the dynamics by exploiting the elegant scheme of Faddeev and
Jackiw \cite{fj} and follow the notation of a recent related work
\cite{duv}.

A generic first order Lagrangian, expressed in the form,
\begin{equation}\label{e1}
L=a_{\alpha}(\eta ) \dot{\eta}^{\alpha}-H(\eta ),
\end{equation}
where $\eta$ denotes phase space variables, leads to the
Euler-Lagrange equations of motion,
\begin{equation}\label{e2}
\omega_{\alpha\beta}\dot
\eta^{\beta}=\partial_{\alpha}H~,~~\omega_{\alpha\beta}=\partial_{\alpha}a_{\beta}-\partial_{\beta}a_{\alpha}.
\end{equation}
where $\omega_{\alpha\beta}$ denotes the symplectic two form. The
inverse of the symplectic matrix is given by
$\omega^{\alpha\beta}$.
\begin{equation}\label{e3}
\omega^{\alpha\beta}\omega_{\beta\gamma}=\delta^{\alpha}_{\gamma}
\end{equation}

For our model, following (\ref{a1}), $\omega^{\alpha\beta}$ is
defined by,
\begin{equation}\label{ee4}
\omega^{\alpha\beta}=
 \left (
\begin{array}{cc}
-\theta(X_iP_j-X_jP_i)  & (\delta_{ij}-\theta P_iP_j) \\
-(\delta_{ij}-\theta P_iP_j) &  \theta \epsilon_{ijk}
\frac{X_k}{X^3}
\end{array}
\right ) .
\end{equation}

The particle dynamics is easily derived in a straightforward way
exploiting (\ref{ee4}). We have,
\begin{equation}\label{e7}
\dot X_i=\{X_i,H\}~ ;~~\dot P_i=\{P_i,H\}.
\end{equation}
Considering a simple form of $H$,
$$H=\frac {P^2}{2m}+V(X),$$ we obtain,
\begin{equation}
\dot X_i=\frac{1}{m}(1-\theta P^2)P_i+\theta ((E.P)X_i-(E.X)P_i),
\label{e8}
\end{equation}
\begin{equation}
\dot P_i=E_i-\theta (E.P)P_i+\frac{\theta}{m}\epsilon
_{ijk}P_j\frac{X_k}{X^3}. \label{e81}
\end{equation}
In the above we have identified $-(\partial V)/(\partial
X_i)\equiv E_i$, the external electric field. We can rewrite the
equations of motion as,
\begin{equation}\label{ee7}
\dot{\vec X} =\frac{\vec P}{m^*}+\theta (\dot {\vec P}\times \vec
L),
\end{equation}

\begin{equation}\label{ee8}
\dot{\vec P}=\vec E-\theta (\vec E. \vec P)\vec P+\frac{\theta
}{mX^2}(\vec P\times \vec S),
\end{equation}
where $m^*=m(1-\theta P^2)^{-1} $ and the spin vector $\vec S$ has
been defined in (\ref{mm1}). It is interesting to note that the
origins of Berry curvature terms in (\ref{ee7}) and (\ref{ee8})
are different: the Snyder form of spatial noncommutativity in
(\ref{aa1}) is responsible for the former, whereas monopole form
of the momentum noncommutativity in (\ref{aa1}) for the latter. We
can also express (\ref{ee7}) as
\begin{equation}\label{eee7}
\dot{\vec X} =\frac{\vec P}{m^*}+\theta (\vec E\times \vec L).
\end{equation}
It is straightforward to iterate  (\ref{e8}) once again so that we
obtain a generalized Lorentz force equation in the following form,
\begin{equation}\label{ee9}
\ddot X_i=\frac{1}{m}(1-\theta P^2)E_i-\frac{3\theta }{m}(E.P)P_i
+\frac{\theta }{m}\epsilon _{ijk}\frac{P_jX_k}{X^3}-\theta
\epsilon _{ijk}(\vec E\times \vec X)_kE_j.
\end{equation}
We will study the  significance of these equations in the
Discussion, Section V, at the end.

\section{The Aharonov-Bohm effect on NC (Snyder) space}
In non-commutative space many interesting quantum mechanical
problems have been studied extensively:  such as hydrogen atom
spectrum in an external magnetic field \cite{h1,h2}, Aharonov-Bohm
(AB) \cite{jab,li}, Aharonov-Casher effects \cite{ac}, to name a
few. However, all the above works have considered a
{\it{constant}} form spatial noncommutativity. In the present work,
for the first time, we consider such effects in the presence of an
{\it{operatorial}} form of noncommutativity. Here we consider a
purely Snyder form of noncommutative space,
\begin{eqnarray}\label{a01}
\left \{X_i, X_j\right\}&=&  -\theta(X_i P_j - X_j P_i),\nonumber\\
\left\{X_i, P_j\right\}&=&\delta_{ij}  -\theta P_i P_j,\nonumber\\
\left\{P_i, P_j\right\}&=& 0,
\end{eqnarray}
that we obtained from (\ref{aa1}) by putting   $\{P_i,P_j\}=0$.

In the commutative Aharonov-Bohm effect, the presence of the flux
produces a shift in the interference pattern. The value of the
flux is such that the position of maxima and minima are
interchanged due to a change of $\pi$ in the phase and vanishes
when magnetic field is quantized. For noncommutative Aharonov-Bohm
effect a velocity dependent extra term in the flux arises even in
the presence of quantized magnetic field \cite{jab}. This could be
experimentally measured. The velocity can be so chosen that  the
phase shift become $2\pi$ or integer multiple of $2\pi$. So this
phase shift might not be observed. The  Aharonov-Bohm effect in
noncommutative case can also be worked out using path integral
formulation \cite{jab}. Electrons moving on a noncommutative plane
in uniform external magnetic and electric field represents usual
motion of electrons in an effective magnetic field. The related AB
phase can be calculated and it yields the same effective magnetic
field \cite{li}. Using non-commutative quantum mechanics
Aharonov-Bohm phase can be obtained on NC phase space  \cite{li}.

  For the NC phase space (\ref{a01}), the variables $X_i,P_j$ can be expressed in terms
  of canonical (Darboux) set of variables $x_i,p_j$:
  \begin{equation} \label{ab1}
X_i=x_i-\theta (x.p) p_i ~;~~    P_i=p_i
\end{equation}
The $x_i,p_j$ satisfy
$$ \{x_i,p_j\}=\delta _{ij}~;~~\{x_i,x_j\}=\{p_i,p_j\}=0.$$

Let $H(X,P)$ be the Hamiltonian operator of the usual quantum
system, then the  Schr\"{o}dinger equation on NC space is written
as
\begin{equation}\label{ab2}
H(X,P)*\psi=E\psi .
\end{equation}

The star product can be changed into the ordinary product by
replacing $H(X,P)$ with $H(x,p)$ \cite{bop}. Thus the Schr\"{o}dinger
equation can be written as,
\begin{equation}\label{ab3}
H(X_i,P_i)\psi=H(x_i-\theta (x.p)p_i ~;~p_i)\psi=E\psi .
\end{equation}

When magnetic field is applied, the Schr\"{o}dinger equation
becomes
\begin{equation}\label{ab4}
H(X_i,P_i,A_i)*\psi=E\psi .
\end{equation}
Now we also need to replace the vector potential $A_i$ with a
phase shift as given by
\begin{equation}\label{ab5}
    {\cal A}_i \rightarrow A_i - \frac{1}{2}\theta ~(x.p)p_j  \partial_j
    A_i .
\end{equation}
So, the Schr\"{o}dinger equation (\ref{ab4}) in the presence of
magnetic field becomes
\begin{equation}\label{ab6}
H(x_i-\theta~(x.p)p_i ~;~p_i~;~A_i - \frac{1}{2}\theta~(x.p) p _j
\partial_j A_i)\psi=E\psi .
\end{equation}
If $\psi_0$ is the solution  of (\ref{ab6}) when $A_i=0$, then the
general solution of (\ref{ab6}) may be given by
\begin{equation}\label{ab7}
    \psi=\psi_0 exp\left[iq\int^x_{x_0} (A_i -
    \frac{1}{2}\theta ~(x.p)p _j  \partial_j A_i) dx_i\right].
\end{equation}
The phase term of (\ref{ab7}) is called the AB phase. In a double
slit experiment if we consider the  charged particle of charge $q$
and mass $m$ to pass through one of the the slits, then the
integral in (\ref{ab7}) runs from the source $x_0$ to the screen
$x$, the interference pattern will depend on the phase difference
of two paths. The total phase shift for the AB effect is
\begin{equation}\label{ab8}
  \Delta \Phi_{AB}=  \delta \Phi_0 + \delta \Phi^{NC}
    = iq\oint A_i dx_i -
   i\frac{q}{2}\theta \oint (mv_j+qA_j) (mv_k+qA_k) x_k \partial_j A_i)
   dx_i ,
\end{equation}
where we use the equation $mv_l=p_l -qA_l$. The line integral runs
from the source through one slit to the screen and returns to the
source through another slit. The first term of (\ref{ab8}) is the
usual AB phase. One of the four $\theta$ dependent term in (\ref{ab8}) is
    $\oint  v_j v_k x_k \partial_j A_i dx_i=
     \vec{\nabla}A_i)(\vec{v}.\vec{x}) dx_i$.
The rest are computed in a similar way.


Adding all the terms we can write the AB phase as
\begin{equation}\label{ab13}
\Delta \Phi_{AB}
    = iq\oint A_i dx_i -
   i\frac{q}{2} \theta \oint \left[(m\vec{v}+q\vec{A}).~\vec{x}\right]
   \left[(m\vec{V}+q\vec{A}).
   \vec{\nabla}A_i\right]dx_i
\end{equation}
Previous results \cite{jab,li} with a {\it{constant}} form of
spatial noncommutativity  are of the form,
\begin{equation}\label{ab14}
\Delta \Phi_{AB}^{NC} \sim i\frac{q}{2}\oint
\left[\vec{\theta}.~(m\vec{v}+q\vec{A}) \times
   \vec{\nabla}A_i\right]dx_i .
\end{equation}
Comparing our result with the above (\ref{ab14}), we find that the
NC correction in our case altogether a different structure and
also there appears an  extra piece $(m\vec{v}+q\vec{A}).\vec{x}$,
which is a consequence of the form of space-space noncommutativity
chosen in our model. We discuss the implications of our result in
the Discussion section.

\section{Discussions:}
Let us summarize and motivate our results. Our aim has been to
demonstrate effective models of interest, (specially in the
context of Condensed Matter physics), can be simulated compactly
in a purely Hamiltonian formulation, developed in a suitable
noncommutative space. The advantage is that one can have a simple
form of Hamiltonian and the complicated responses of the system
are induced by the noncommutative structure of spacetime. To be
specific, in the present work in Section III, we have shown that
effective {\it{spin}}-contributions are generated in our model,
only from the symplectic structure with no explicit spin-term as
such.

It was shown in \cite{berard} that the anomalous velocity term
related to spin Hall effect has a natural interpretation in terms
of a noncommutative space. We have shown that this result can be
generalized in various possible ways: One can have Berry curvature
terms both in coordinate as well as in momentum space and the
singularity structure of the Berry potential, (not shown here) ,
can be more complicated than that of a monopole. This will become
apparent if one inverts $\omega ^{\alpha\beta}$ in (\ref{ee4}) to
recover the Berry potential.

A novel form of anomalous velocity term has been derived in
(\ref{e8}). From the equivalent form (\ref{ee7}) we infer that
there will be a deviation in the particle trajectory in the
presence of an electric field \cite{horvathy}. On the other hand,
in (\ref{ee8}) we have an explicitly spin dependent term in the
expression of $\dot{\vec P}$ with a coordinate space singularity.
Once again, in the alternative force equation in (\ref{ee9}), the
leading term in low energy $\theta \frac{\vec P\times \vec
S}{X^2}$ reminds us of models with the Rashba type of interactions
\cite{nik}. Hence, these effects can be relevant for the studies
in \cite{pur,kawa}.

Now we come to the results obtained in Section IV and their
implications. As we have mentioned in Section IV, we consider the
Snyder noncomutative space, as given in (\ref{a01}). As we have
pointed out in Section IV, in the present case, the $\theta
$-contribution in the AB phase, (derived for the constant
noncommutative case \cite{jab,li}), gets multiplied by  a
dynamical factor. This leads to some interesting consequences. As
in previous cases \cite{jab}, we can also derive a bound on
$\theta $ pertaining to experimental observations. We compute
$\gamma $, the ratio of the AB phases appearing in the normal case
and noncommutative case:
\begin{equation}\label{d1}
\gamma\equiv \frac{\Delta \phi ^{NC}}{\Delta \phi}\sim
(\frac{\theta }{R\lambda _e}\frac{v}{c})(\frac{R}{\lambda
_e}\frac{v}{c})\approx \frac{\theta }{\lambda
_e^2}(\frac{v}{c})^2.
\end{equation}
In (\ref{d1}), $\Delta \phi ^{NC}$ corresponds to the $\theta
$-contribution in (\ref{ab13}) and $\Delta \phi $ refers to the
$\theta =0$ commutative case, $R$ denotes the electron radius in
the experimental setup and $\lambda _e$ is the Compton wavelength
of the electron. Interestingly, in the present case, the extra
dynamical factor cancels $R$ in $\gamma $ and reproduces the bound
in the $R$-independent form:
\begin{equation}\label{d2}
\sqrt \theta < (\frac{v}{c})^{-1}\lambda _e .
\end{equation}
This is distinct from the previously obtained expressions
\cite{jab} but the bound is much lowered than that of \cite{jab}.

Finally, we would like to make a remark on the effect a generic
noncommutative space can have in the study of {\it{inequivalent}}
quantization in a non-simply connected manifold \cite{path}. It is
well known that AB effect is a prototype example of a multiply
connected domain since the region of the solenoid that carries the
magnetic flux is inaccessible to the charged particle. This leads
to a punctured manifold $Q=\mathbb{R}^2-\delta $, ($\delta $
denoting the solenoidal area), with a non-trivial first homotopy
group $\Pi _1(Q)=Z$. One can still work in the trivial homotopy
sector, but this requires additional topological terms in the
action. They clearly show up in the path-integral quantization of
the system. These issues have been extensively studied in
\cite{path}, for the normal (commutative) spacetime. As it has
been established here and before \cite{jab,li}, that
noncommutative nature of spacetime generates additional
contributions in the AB phase, clearly this will directly affect
the above mentioned quantization programme. From the study of the
modified quantization conditions, it might be possible to set an
independent bound on $\theta $. We intend to study this aspect in
future. \vskip .5cm

{\it {Acknowledgements}}: We would like to thank Prof.P.Horvathy
for discussions and Prof.G.Tatara and Prf.D.Xiao for
correspondences. Also we thank the Referees for the constructive
comments.


\begin{thebibliography}{99}
\bibitem{dirac} P.A.M.Dirac, Proc.Roy.Soc.London 133(1931)60 ;\\
R.Jackiw, Phys.Rev.Lett. 54(1985)159; For a different perspective
and more related to our work, see R.Jackiw, Int.J.Mod.Phys.
A19S1(2004)137.
\bibitem{thooft}G. t'Hooft, Nucl.Phys. B79 (1974) 276; A.M.Polyakov, JETP Lett. 20 (1974)194.
\bibitem{mono1}M.Onoda and N.Nagaosa, J.Phys.Soc.Jpn. 71(2002) 19; Z.Fang et al., Science 302 (2003) 92.
\bibitem{lutt}R.Karplus and J.M.Luttinger, Phys.Rev. 95(1954)1154.
\bibitem{berard}A.Berard and H.Mohrbach. Phys.Rev. D69 95
(2004) 127701.
\bibitem{nc}
H. S. Snyder, Phys. Rev. {\bf 71}(1947) 38, N.Seiberg and E.Witten, JHEP 9909(1999)032,
R.J.Szabo, Phy. Rep. 378 (2003)207,
  J. Lukierski, H. Ruegg, W.J. Zakrzewski, Annals Phys. 243 (1995)90,
  M.Dimitrijevic, L.Jonke, L.Moeller, E.Tsouchnika, J.Wess and M.Wohlgenannt,  Czech.J.Phys. 54 (2004)1243;
 S.Doplicher, K.Fredenhagen and J.E.Roberts, Phys.Lett. B331 (1994)
39; 
 S.Ghosh and P.Pal, Phys.Lett. B633 (2006)397
  Phys.Lett.B 618 (2005)243 
 S.Ghosh, Phys.Lett.B 623 (2005)251 
\bibitem{niu}
G.Sundaram and Q.Niu,
Phys.Rev.B B59(1999)14915; F.D.M.Haldane, Phys.Rev.Lett. 93 (2004)
206602 cond-mat/0408417; R.Shindou and K.-I Imura,
Nucl. Phys. B720[FS](2005) 399 ; C.Zhang, A.M.Dudarev and Q.Niu,
cond-mat/0507125.
\bibitem{pur}P.Pureur et.al., cond-mat/0501482.
\bibitem{kawa} G.Tatara and H.Kawamura, J.Phys.Soc.Jpn 71 (2002) 2613; 
 \bibitem{gen1} M.Onoda, G.Tatara and N.Nagaosa, J.Phys.Soc.Jpn. 73 (2002) 2624; H.Kawamura, Phys.Rev.Lett. 90 (2003) 047202,
G.Tatara (correspondence).
\bibitem{lages}  A. Berard, J. Lages and H. Mohrbach,  Eur.Phys.J. C35 (2004)
373-381; S.Ghosh, Phys.Lett. B638 (2006)350 
\bibitem{fj}L.Faddeev and R.Jackiw, Phys.Rev.Lett. 60 (1988) 1692.
\bibitem{duv}C.Duval et al., Mod.Phys.Lett B20 (2006) 373 ;
Z.Horvath, P.A.Horvathy and L.Martina, cond-mat/0511099.
\bibitem{h1}C. Duval and P. A. Horváthy, Phys.Lett. B, Volume 479, (2000)
284; V. P. Nair, Phys. Lett. {\bf B 505} (2001) 249 ; V. P. Nair
and A. P. Polychronakos, Phys. Lett. {\bf B 505} (2001) 267 .
\bibitem{h2} M.Chaichian, M.M.Sheikh-Jabbari and A.Tureanu, Phys. Rev. Lett. {\bf 86} (2001) 2716 ; M.Chaichian, A. Demichev, P.Presnajder, M.M.Sheikh-Jabbari and A.Tureanu, Nucl. Phys. {\bf B 611} (2001) 383 .
\bibitem{jab} M.Chaichian, P.Presnajder, M.M.Sheikh-Jabbari and A.Tureanu, Phys.Lett. B527 (2002)
149; H. Falomir, J. Gamboa, M. Loeve, F. Mendez and J. C.
Rojas, Phys. Rev. {\bf D 66} (2002) 045018 .
\bibitem{li}K.Li and S.Dulat, hep-th/0508193.
\bibitem{ac} B. Mirza and M. Zarei, Eur. Phys. J. {\bf C 32} (2004) 583 .
\bibitem{bop} T. Curtright, D. Fairlie and C. Zachos, Phys. Rev. D58 (1998) 25002;
L. Mezincescu, hep-th/0007046.
\bibitem{horvathy} P.A.Horvathy, hep-th/0602133
\bibitem{nik} B.Nikolic, L.P.Zarbo and S.Welack, Phys.Rev. B72 (2005)
075335 . We thank D.Xiao for this.
\bibitem{path}See for example, L. Schulman, {\it{Techniques and applications of path integrals}}, Wiley
and Sons, N. Y. 1981; P.A.Horvathy, G. Morandi and
E.C.G.Sudarshan, Il Nuovo Cimento D11, p. 201 (1989); H. Kleinert,
{\it{Path Integrals in Quantum Mechanics, Statistics, and Polymer
Physics}}, 2nd Ed. World Scientific (1995).

\end{thebibliography}
\end{document}